\documentclass[a4paper,fleqn]{revtex4}
\usepackage{graphicx,rotating}
\begin{document}
\title{High Resolution Rydberg Spectroscopy of ultracold Rubidium Atoms}
\author{Axel Grabowski, Rolf Heidemann, Robert L\"{o}w, J\"{u}rgen Stuhler, and Tilman Pfau}
\affiliation{5.Physikalisches Institut, Universit\"{a}t Stuttgart,
Pfaffenwaldring 57, 70550 Stuttgart, Germany}
\date{\today}
\begin{abstract}
We present experiments on two-photon excitation of ${\rm ^{87}}$Rb
atoms to Rydberg states. For this purpose, two continuous-wave
(cw)-laser systems for both 780 nm and 480 nm have been set up.
These systems are optimized to a small linewidth (well below 1
MHz) to get both an efficient excitation process and good
spectroscopic resolution. To test the performance of our laser
system, we investigated the Stark splitting of Rydberg states. For
n=40 we were able to see the hyperfine levels splitting in the
electrical field for different finestructure states. To show the
ability of spatially selective excitation to Rydberg states, we
excited rubidium atoms in an electrical field gradient and
investigated both linewidths and lineshifts. Furthermore we were
able to excite the atoms selectively from the two hyperfine ground
states to Rydberg states. Finally, we investigated the
Autler-Townes splitting of the 5S$_{1/2}$$\rightarrow$5P$_{3/2}$
transition via a Rydberg state to determine the Rabi frequency of
this excitation step.
\end{abstract}
\maketitle                   
\section{Introduction}
During the last decades, a lot of studies have been performed on
Rydberg atoms \cite{Gallagher}. These experiments made use of
atomic beams emitted from a thermal source. In the last years,
methods of laser cooling \cite{Metcalf} of atoms opened new ways
to perform experiments on Rydberg states using atom samples at
temperatures in the 100 ${\rm \mu}$K range. Such a system of
ultracold Rydberg atoms is known as "'frozen Rydberg gas "'
\cite{Anderson, Mourachko}. The name is correlated to the fact,
that there is nearly no thermal motion of the atoms during the
time constant of the experiment. Typically, in the time for one
experiment of ${\rm \sim 10 \mu }$s, an atom moves about ${\rm 1
\mu }$m which is smaller than the mean interatomic distance. An
interesting property of Rydberg atoms are the long-range
dipole-dipole and van der Waals interactions. First investigations
of the interaction among the atoms in such "'frozen Rydberg"'
systems have started recently \cite{Singer, Tong, Carroll,
Afrousheh, Li}. In these experiments, the interaction led to an
increasing suppression of excitation as a function of increasing
Rydberg atom density. This suppression mechanism is called van der
Waals blockade, because the van der Waals interaction among the
atoms prohibits further excitation by shifting the energy levels
of neighboring atoms out of resonance with respect to the
excitation lasers.\\
Rydberg atoms or mesoscopic ensembles of them are discussed in the
context of quantum information processing \cite{Jaksch, Lukin} as
conditional logic gates based on the strong interaction between
the atoms. To use Rydberg atoms in quantum computing, there are
several tasks which have to be achieved first. In the proposal
given by Jaksch et al. \cite{Jaksch}, there must be a possibility
to arrange the atoms spacially in a defined way. This can either
be done in optical \cite{Grimm} or magnetic lattices
\cite{Grabowski}, where ground state atoms are captured. The
quantum information is hereby stored in the hyperfine ground
state of the atoms.\\
In this paper, we present experiments on the excitation of cold
${\rm ^{87}}$Rb atoms to Rydberg states. The atoms are first
prepared in a magneto-optical trap (MOT) \cite{Metcalf} and then
excited in a two-photon two-color excitation step to a Rydberg
state. For this purpose, a narrow linewidth cw-laser system for
both wavelengths has been set up. Using this system, we
investigated Rydberg excitation in the proximity of the n=40
state. With the high spectroscopic resolution of our laser system,
we were able to investigate the Stark splitting of the Rydberg
states and could even resolve the state dependent hyperfine
Stark splitting for different J-states.\\
For quantum computing purposes with Rydberg atoms \cite{Jaksch}
one has to be able to address the atoms individually in space (for
example in an optical lattice). Due to the fact that the quantum
information is stored in the hyperfine ground states, it is
necessary to be able to excite the atoms from a defined ground
state (in ${\rm ^{87}}$Rb, F=1 and F=2) to the Rydberg state. For
every quantum computing scheme, it is essential to conserve
coherence during the operations. In case of using Rydberg atoms,
this means that a coherent excitation of the atom to Rydberg
states is needed. To get to know the time constants of the
experimental cycle, we measured the Rabi frequency of the lower
transition of the two photon excitation process. For this purpose,
we investigated the Autler Townes splitting of the
5S$_{1/2}$$\rightarrow$5P$_{3/2}$ transition. This is done by
probing the splitting of this splitted state  with a narrow band
excitation to a Rydberg state \cite{Teo}. \\
This paper is organized as follows: In section 2, we describe the
experimental setup, especially the laser system for the Rydberg
excitation and the detection of the atoms. In section 3, we give
an example for the spectroscopic resolution and the overall
stability of the system by measuring a Stark map in the vicinity
of the n=40 manifold. Section 4 presents experiments on the state-
and spatially selective excitation. Beyond this, we report in
section 5 on the measurements of the Rabi frequency on the
5S$_{1/2}$$\rightarrow$5P$_{3/2}$ transition by observing the
Autler-Townes splitting of this transition.
\section{Experimental setup and cold atom preparation}
\subsection{Vacuum system and magneto optical trap (MOT)}
\begin{figure*}[h]
\includegraphics[width=14cm]{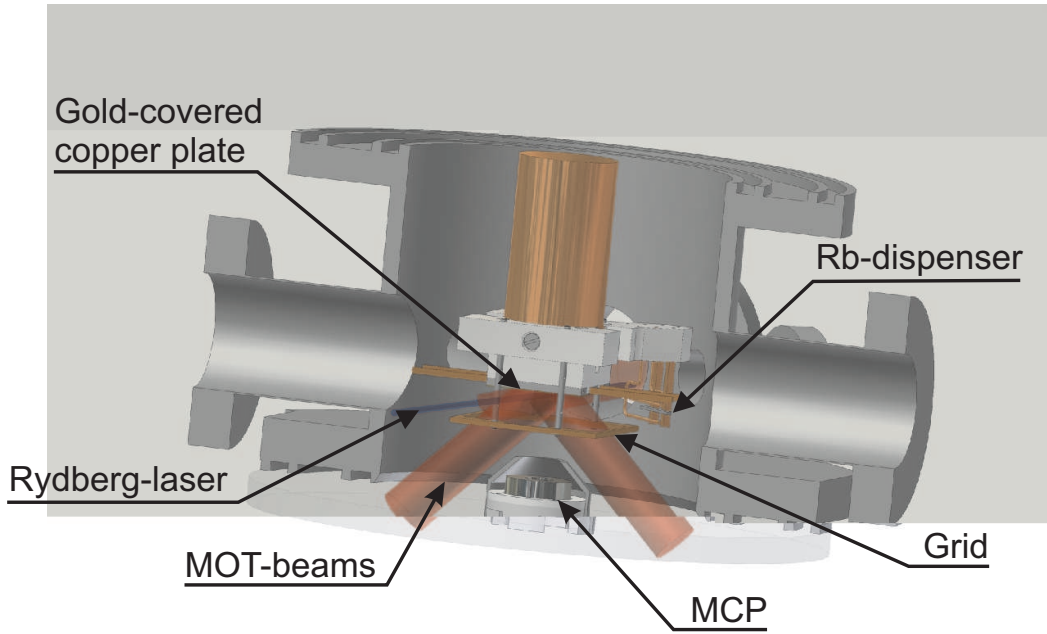}
\caption{Setup of the vacuum system with field plates and micro
channel plate (MCP). Furthermore the laser beams for both MOT
(red) and Rydberg excitation are drawn (blue beam in front). The
gold covered copper plate and the copper grid are 10 mm apart from
each other (middle of the picture) surrounding the MOT. The
dispenser in the background is shielded by a cage. The MCP is
situated in the middle of the vacuum chamber on the lower window.}
\label{fig:1}
\end{figure*}

The experiments presented here are performed inside a
vacuum-chamber as shown in fig. \ref{fig:1}. The chamber consists
of a steel tube with an inner diameter of 10 cm. In the radial
direction, optical access is provided on three axes (4x CF-36
flanges with windows and 2 x CF-16 flanges with windows). The
bottom of the chamber is sealed with a large glass window of 175
mm in diameter. The MOT is built up as a reflection MOT
\cite{Reichel2} where two of the MOT beams are reflected from a
reflective surface inside the vacuum chamber. The other MOT beams
use one of the accessible axes of the chamber. One other axes of
the chamber is used for fluorescence imaging of the atomic cloud.
The magnetic quadrupole field for the MOT is produced by the
current through a wire behind the gold covered copper plate inside
the vacuum chamber superimposed with a homogeneous magnetic field
produced by Helmholtz coils outside the chamber
(U-type quadrupole) \cite{Reichel}.\\
The rubidium for the MOT is provided by a dispenser (SEAS Getters)
situated close to the atomic cloud and well shielded by a mesh
cage to prohibit any disturbances in the electrical field inside
the vacuum chamber. The pressure inside the vacuum chamber which
is pumped by both a titanium sublimation pump and an ion pump is
in the range of $10^{-10}$ mbar. \\
Due to the fact, that Rydberg atoms are very sensitive to
electrical fields \cite{Gallagher}, all sources for electrical
stray fields inside the vacuum have to be shielded as good as
possible. To control the electrical field, we placed field plates
inside the vacuum chamber around the MOT region. Here, we used the
gold covered Cu plate which is likewise used for the reflection
MOT as one field plate. A copper mesh (SPI Fine Grid Mesh,
02199C-AG), sitting 1 cm below the copper plate forms the second
field plate. The mesh has a high optical transmission of 85 \% and
the size of the holes in the grid is only 234 $\mu$m. These field
plates enable us to apply a homogeneous electrical field and
shield the MOT region from electric field gradients. By applying a
voltage to these plates, we can generate an electrical field and
are able to investigate the Stark interaction of the Rydberg atoms
with it.\\
\begin{figure*}[h]
\includegraphics[height=6cm]{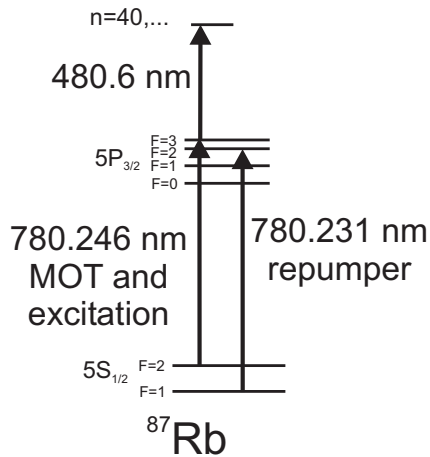}
\caption{Level scheme of ${\rm ^{87}}$Rb for cooling, trapping and
Rydberg excitation.} \label{fig:2}
\end{figure*}
For the magneto optical trap, we use grating stabilized diode
lasers in the Littrow configuration \cite{Ricci}. The MOT-laser is
quasi-resonant to the 5S$_{1/2}$(F=2) $\rightarrow$5P$_{3/2}$(F=3)
transition ($\lambda$=780.248 nm, see fig. \ref{fig:2}). It is
stabilized via polarization spectroscopy on a crossover transition
in the Rubidium spectrum \cite{Demtroeder}, 133 MHz below the MOT
transition. The laser light is afterwards amplified in a "'slave"'
diode and then frequency shifted by an accusto-optical modulator
(AOM). The frequency is usually detuned by -2$\Gamma$ with respect
to the 5S$_{1/2}$(F=2) $\rightarrow$ 5P$_{3/2}$ (F=3) transition.
We end up with a total power of about 30 mW in the 3 MOT beams
(all of them are retro-reflected). As second laser for the MOT,
another grating stabilized diode laser is also locked on a
crossover signal. This laser is shifted with an AOM in single pass
to resonance with the 5S$_{1/2}$(F=1)$\rightarrow$5P$_{3/2}$(F=2)
transition ($\lambda$=780.231 nm, see fig. \ref{fig:2}). This
second so-called repumper laser is needed because MOT atoms can be
pumped to the lower hyperfine ground state by off-resonant
excitation and are yet lost out of the MOT cycle. The repumper
pumps these otherwise lost atoms back. It is superimposed with all
the beams of the MOT laser. The AOMs in both of the beams can also
be used for fast switching.\\
For the detection of the MOT atoms, a CCD camera outside the
vacuum chamber is used to record both fluorescence and absorption
images. The cloud is imaged with a 1:1 telescope to the camera.
For the experiments presented here, only fluorescence pictures are
taken. The number of atoms in the MOT can be calculated from these
fluorescence pictures. The temperature of the atoms in the
MOT can be determined by the time-of-flight expansion of the cloud.\\
We usually operate the Rubidium dispenser at low currents of 4-4.5
A to prevent contamination of the vacuum chamber during the
experiments. This leads to MOT atom numbers of $10^6$ atoms at a
density of ${\rm 10^{10}}$ atoms/cm$^3$.  The temperature of the
MOT is about 100-400 $\mu$K.\\
\subsection{Rydberg laser system and Rydberg excitation }
As shown in fig. \ref{fig:2}, two lasers with wavelengths of 780
nm and 480 nm are needed for the two-photon Rydberg excitation.
For this purpose, we built up a laser system consisting of three
diode lasers. For the 480 nm laser system the problem arises, that
one wants to be able to stabilize the laser to every possible
optical Rydberg transition between n=30 and the ionization
threshold. For this reason, we decided to stabilize a master laser
to a reference resonator. This 30 cm long reference cavity is
stabilized in length by means of a highly stable laser system.
With this setup we gain the possibility to lock the laser on every
resonator fringe, at a separation of 125 MHz. The fringe to lock
on is first found with a wavemeter (Advantest TQ~8325, 100 MHz
resolution) and afterwards by scanning the laser around the
Rydberg resonance with an AOM. The laser system is built up as
master-slave diode laser system at a wavelength around 960 nm. The
master laser is as described before a cavity stabilized diode
laser, which is locked to the reference cavity. The laser beam is
afterwards frequency shifted with a 400 MHz AOM in double pass
configuration, which allows us to vary the frequency of more than
250 MHz. After passing the AOM the beam seeds a 300 mW slave
diode. If the master laser is locked onto the reference cavity it
is possible to change the slave frequency by shifting the
frequency of this AOM. With this set up, it is possible to
generate every frequency in between two resonator fringes.
Afterwards this 960 nm laser is frequency doubled in a home built
doubling resonator by second harmonic generation in a KNbO$_3$
crystal, which yields a laser power of up to 20 mW at 480 nm after
the cavity. This blue beam passes afterwards a 200 MHz AOM in
single pass, which is used to regulate
the power and to switch the beam.\\
With our laser system it is possible to scan the master-slave
system unlocked to the cavity mode hop free about 6 GHz
continuously (respectively frequency doubled 12 GHz). This scan
range is limited by the available voltage range we apply to the
piezo actuator in the master laser. By locking the laser to the
reference cavity, the laser frequency is basically fixed but the
frequency of the slave laser can be adjusted using the AOM between
master and slave laser. This way, we can scan the blue laser
frequency about 500 MHz with high stability and small linewidth.\\
For the 780 nm wavelength, an external cavity stabilized diode
laser is used, locked to a rubidium cell via polarization
spectroscopy. This laser has two applications. It is both used as
a reference for stabilizing the length of our reference cavity and
as first laser of the two-photon transition. This laser beam
passes a 200 MHz double pass AOM which enables us to switch the
laser fast, attenuate the beam and vary the laser frequency by
more than 100 MHz. Depending on the line in the polarization
spectroscopy spectrum we lock the laser to, we are able to work on
the 5S$_{1/2}$(F=2)$\rightarrow$5P$_{3/2}$(F=3) resonance or up to
500 MHz blue detuned with respect to this transition.\\
Both of the laser beams are guided to the experiment by two
singlemode optical fibers. After the optical fiber, we have
typically 2.5 mW for the blue 480 nm beam and for the red 780 nm a
power of about 7 mW. The red laser is collimated (${\rm 1/e^2}$
diameter of 0.95 mm) and shown onto the atomic cloud. The beam
diameter is therefore larger than the MOT diameter. The blue laser
is focused into the MOT cloud, and has a ${\rm 1/e^2}$ diameter of
20 $\mu$m. Its Rayleigh range of 3 mm is much larger than the size
of the cloud. The overlapped pair of beams is irradiated inside
the chamber. The beams are slightly tilted with an angle of about
$10^{\circ}$ with respect to the main
axes of the chamber.\\
Due to the selection rules for the two-photon excitation of atoms
from the 5S$_{1/2}$ via the 5P$_{3/2}$ state to a Rydberg state
only excitations to nS and nD states are dipole allowed
transitions. However, even for small electrical fields the
selection rules are no longer strict. Excitations to usually
dipole transition forbidden states become possible because the
target Rydberg states are no longer pure states but a mixture of
different unperturbed states. Therefore, it is possible to excite
atoms to P or even $l>$3 states in an non-vanishing electrical
field. The $l>$3 states are called hydrogen-like states because
the quantum defect for these states is negligible and they can be
seen as pure hydrogen states showing a linear Stark effect with a
permanent dipole moment. In contrast, the states with $l<$3
sustain for small electrical fields a quadratic Stark effect.
\subsection{Detection of the Rydberg atoms}
For the detection of the Rydberg atoms, we use the way of
field-ionizing the Rydberg atoms by an electrical field pulse and
subsequent detection of the emerging ions with a micro channel
plate (MCP). For the field ionization, the electrical field is
switched on by a high speed, high voltage MOSFET switch (Type:
Behlke HTS-6103 GSM). The typical rise time of this switch is 60
ns. We slow down the switching by a $2^{\rm nd}$ order low pass
filter. This yields a rise time of the electrical field of 55
$\mu$s, which allows us to probe both ions and field-ionized
Rydberg atoms selectively at different times. The ions are
accelerated towards the MCP by the electrical field and are
detected. The MCP signal is amplified in a home built
I-U-converter circuit. The readout of the data is done by a
computer with a digitizer card. The maximum sampling rate of the
analog electronics and the digitizer card is 100 ns.\\
For the calibration of the MCP, we directly photo-ionize the atoms
from the MOT in an electrical field above the ionization
threshold. This leads to a loss of atoms in the MOT due to
ionization. After 100 ms the MOT atom number is decreased to a new
steady state population which is related to the loading rate of
the MOT and the loss rates. Measuring the reduced steady-state MOT
atom number with the CCD-camera and recording simultaneously the
number of produced ions with the MCP, we can calibrate the MCP
efficiency. With this calibration, we are now able to calculate atom
numbers from the MCP signal.\\
\subsection{Excitation sequence}
\begin{figure*}[h]
\includegraphics[height=7cm]{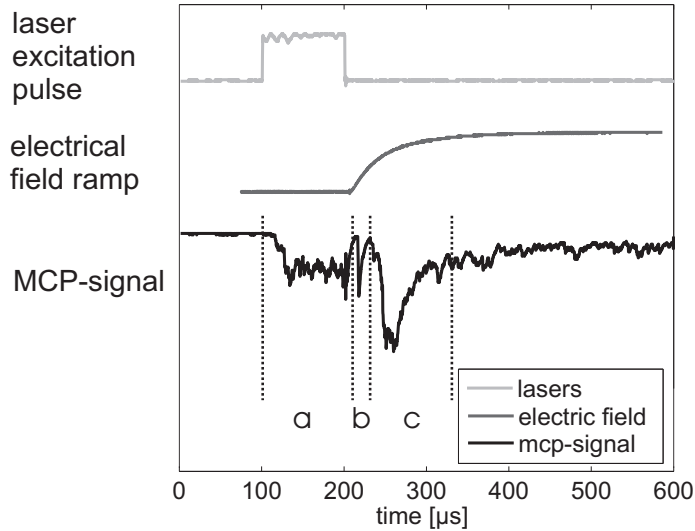}
\caption{Excitation sequence of the Rydberg atoms, field
ionization pulse and corresponding MCP signal (see text). }
\label{fig:3}
\end{figure*}

During the Rydberg excitation all, the MOT and repumper beams are
switched off with the AOMs. After 100~$\mu$s expansion time, the
two collinear beams for Rydberg excitation are switched on
simultaneously for a variable time between 100~ns and 1~ms. With a
variable delay after the excitation pulse a voltage of 300~V is
applied to the gold covered copper plate for field ionization of
the Rydberg atoms. At the same time the copper grid is at
ground potential, which yields an electrical field of 300~V/cm.\\
In fig. \ref{fig:3} an example for the excitation sequence is
shown. Here the excitation pulse is 100 $\mu$s long. Already
during the excitation pulse without an electrical field a signal
on the MCP is seen (a). This signal is attributed to
Rydberg-Rydberg collisions which lead to an ionization process
among the Rydberg atoms \cite{Robinson}. Immediately after the
excitation pulse is switched off, the electrical field is ramped
up. At the beginning of this ramp a first bunch of ions is
detected by the MCP (b). We assign this to ions that first
remained in the atom cloud and are now pulled out by the rising
electrical field of the ionization pulse. About 40 $\mu$s later, a
broad peak appears on the MCP signal (c), which is due to the
field-ionized Rydberg atoms. The time until they reach the MCP is
determined by the rising time of the electrical field. Only when
the electrical field is high enough, the atoms are ionized and the
ions are then accelerated towards the MCP. This peak is broadened
due to l-changing collisions \cite{Walz}. In the data shown in the
next sections (see e.g. fig. \ref{fig:4}), every data point
consists of such a measurement as in fig. \ref{fig:3}. All the
data of the single measurements are collected and stored.
Afterwards the ion signal is integrated to get together with the
calibration factor an absolute Rydberg atom number. The typical
repetition rate of the experiment is 30 ms.
\section{Spectroscopy of Rydberg states, F splitting of the Rydberg states }
\begin{figure*}[h]
\includegraphics[width=\linewidth]{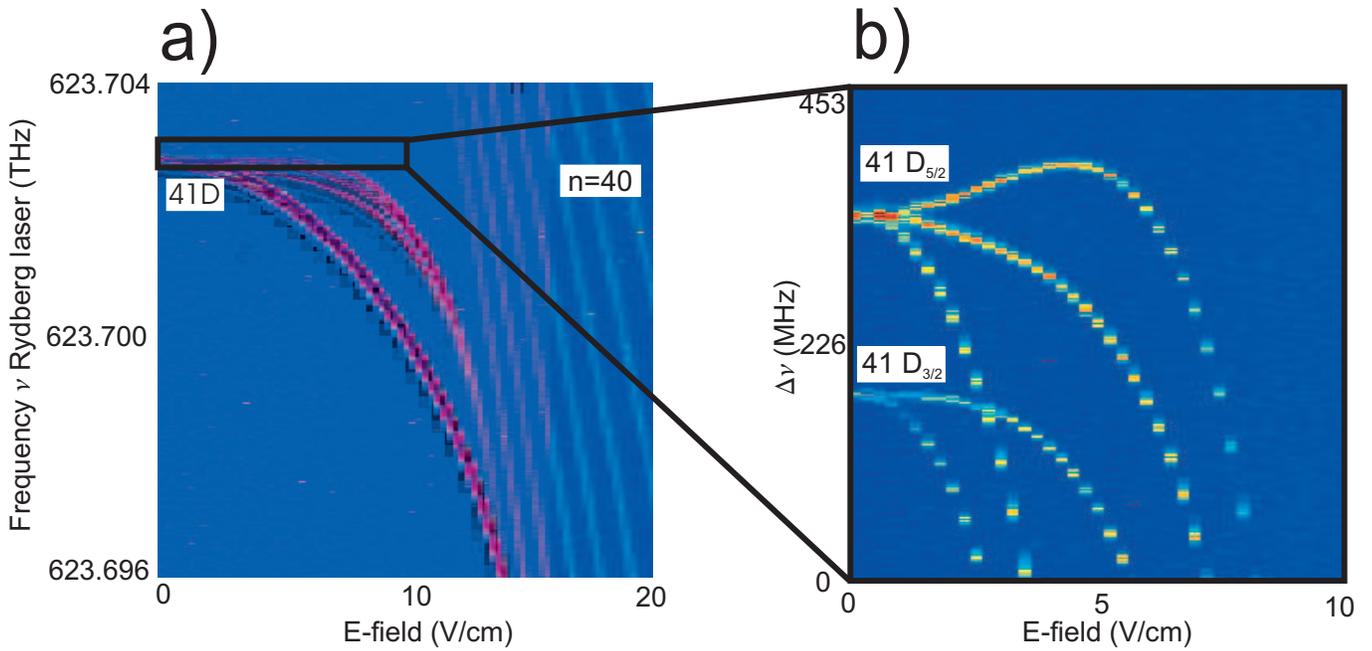}
\caption{Example for the spectroscopic resolution of the laser
system. On the left hand side a Stark map scan is shown where the
electrical field and the blue laser frequency is changed. On the
right side is a scan of a smaller region with a higher resolution,
revealing the F-Splitting of the Rydberg states in the electrical
field.} \label{fig:4}
\end{figure*}

Figure \ref{fig:4} shows a high resolution spectrum of the 41D
doublet. For this experiment, we held the frequency of the red
laser constant on resonance of the of the 5S$_{1/2}$(F=2)
$\rightarrow$5P$_{3/2}$(F=3) transition and changed both the
frequency of the blue laser and the electrical field during
excitation. In fig. \ref{fig:4} a) the spectrum is taken with the
unstabilized blue laser while in b) a spectrum obtained by
scanning the locked laser system is shown. The electrical field
applied across the interaction region during the excitation
sequence is changed between 0 V/cm and 20 V/cm (a), respectively
10 V/cm (b). The color of the pictures indicates the signal
strength (blue-low to red-high signal).\\
On the right hand side in fig. \ref{fig:4} a) both the 41 D
doublet and the manifold of the n=40 states can be found. As
expected the intensity of the n=40 lines is much weaker than the
one of the D-lines, because these are $l>$3 states. Because some
shift in the Rydberg lines cannot be excluded, we improved the
stability by locking the laser onto the reference cavity. In fig.
\ref{fig:4} b) the measurement is done by changing the frequency
of the AOM between master and slave laser. Thus we were able to
scan the laser frequency with a stability of better than 1 MHz.
So the splitting of the F-substates in the electrical field can be
observed. The 41D$_{5/2}$-level has the hyperfine substates
F=1,2,3,4, the 41D$_{3/2}$ the F-states F=0,1,2,3 ($^{87}$Rb has a
nuclear spin of I=$3/2$). We start the excitation from the
5S$_{1/2}$, F=2 state via the 5P$_{3/2}$, F=3 state. Therefore we
can excite the atom in two different 41D$_{3/2}$-, respectively
three different 41D$_{5/2}$- states as can be seen in the picture.
The additional shift of the 5S$_{1/2}$ and 5P$_{3/2}$ states is in
the Stark maps negligible, because the polarizability of
$^{87}$Rb in these low lying states is very small.\\
Beyond this, we measured the n=40 manifold in the environment of
the zero field crossing. From these data we could estimate the
residual electrical field components to be less than  1 ${\rm
V/cm}$. The accuracy of the measurement was only limited by the
small zero field excitation rate.\\
For small detunings and on resonance of the red transition, the
linewidths of the entire transition we observe are basically
determined by the linewidth of the lower transition. For large
detunings from the 5P$_{3/2}$ state, the linewidth becomes
smaller, because the intermediate state is adiabatically
eliminated. The smallest linewidth we have measured so far was 2.2
MHz, although the laser linewidth is well below 1 MHz, determined
from the laser stabilization signal. We attribute the residual
line broadening to remaining magnetic fields of the MOT. The MOT
has a magnetic field gradient of about 16 G/cm. The shift of the
atoms in the S$_{1/2}$ states is 0.7 MHz/G between two neighboring
m$_F$ substates (m$_F$=-2,..,2). At a MOT diameter of 500 $\mu$m
this yields a linewidth due to magnetic field broadening of 2.24
MHz which is in excellent agreement with our measurements.
\section{Spatial and state selective addressing of Rydberg states }
\subsection{Spatial selective Rydberg excitation}
\begin{figure*}[h]
\includegraphics[height=7cm]{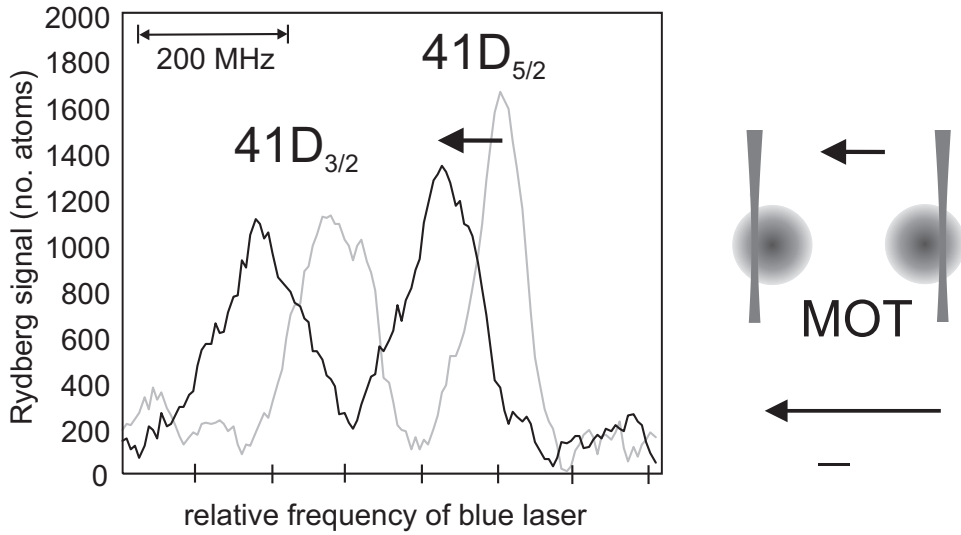}
\caption{Spatial dependent addressing of atoms of the cold cloud
by using an electrical field gradient. The spacial addressing was
done by moving the laser across the MOT. The frequency of the
Rydberg lines is shifted according to the electrical field
gradient.} \label{fig:5}
\end{figure*}

For quantum computing approaches with Rydberg atoms according to
the proposal of D. Jaksch et al. \cite{Jaksch} it is necessary to
address the atoms individually at different positions in space.
One possibility to do this is addressing via the Stark shift in
electrical field gradients which links the resonance frequency of
the atoms to their position. Here we present a demonstration
experiment with Rydberg atoms in an electrical field gradient. For
this purpose, we applied an electrical field gradient across our
MOT and varied the position of the excitation lasers along this
gradient. In fig. \ref{fig:5} the 41D$_{3/2}$ and 41D$_{5/2}$
resonance lines are shown. The shift of the excitation beams by
the diameter of the MOT cloud yields a line shift of 98 MHz and 75
MHz which is already larger than the linewidth of the broadened
transitions. This broadening is due to electrical field gradient
across the MOT. This is because the gradient we applied is not
homogeneous in one direction, but also varies in the orthogonal
direction. So each individual atom along the Rydberg excitation
beam in the MOT is exposed to a different electrical field which
leads to an inhomogeneous broadening. The measured linewidth of
the two lines is 81 MHz, respectively 57 MHz which is much larger
than the linewidth of the 5S$_{1/2}$(F=2)$\rightarrow$5P$_{3/2}$(F=3)
transition.\\
The origin for the different linewidths and shifts of the two
transitions is the different slope of the two transitions in the
electrical field as can be seen in fig. \ref{fig:4}. From the
slope of this state in an electrical field and from the size of
the cloud, the gradient of the electrical field in the direction
of the movement of the laser can be estimated to be 18.8 V/cm$^2$.
In future experiments, we will be able to apply a homogeneous
electric field gradient which will shift the lines without the
line broadening.
\subsection{Hyperfine selective Rydberg excitation}
\begin{figure*}[h]
\includegraphics[width=\linewidth]{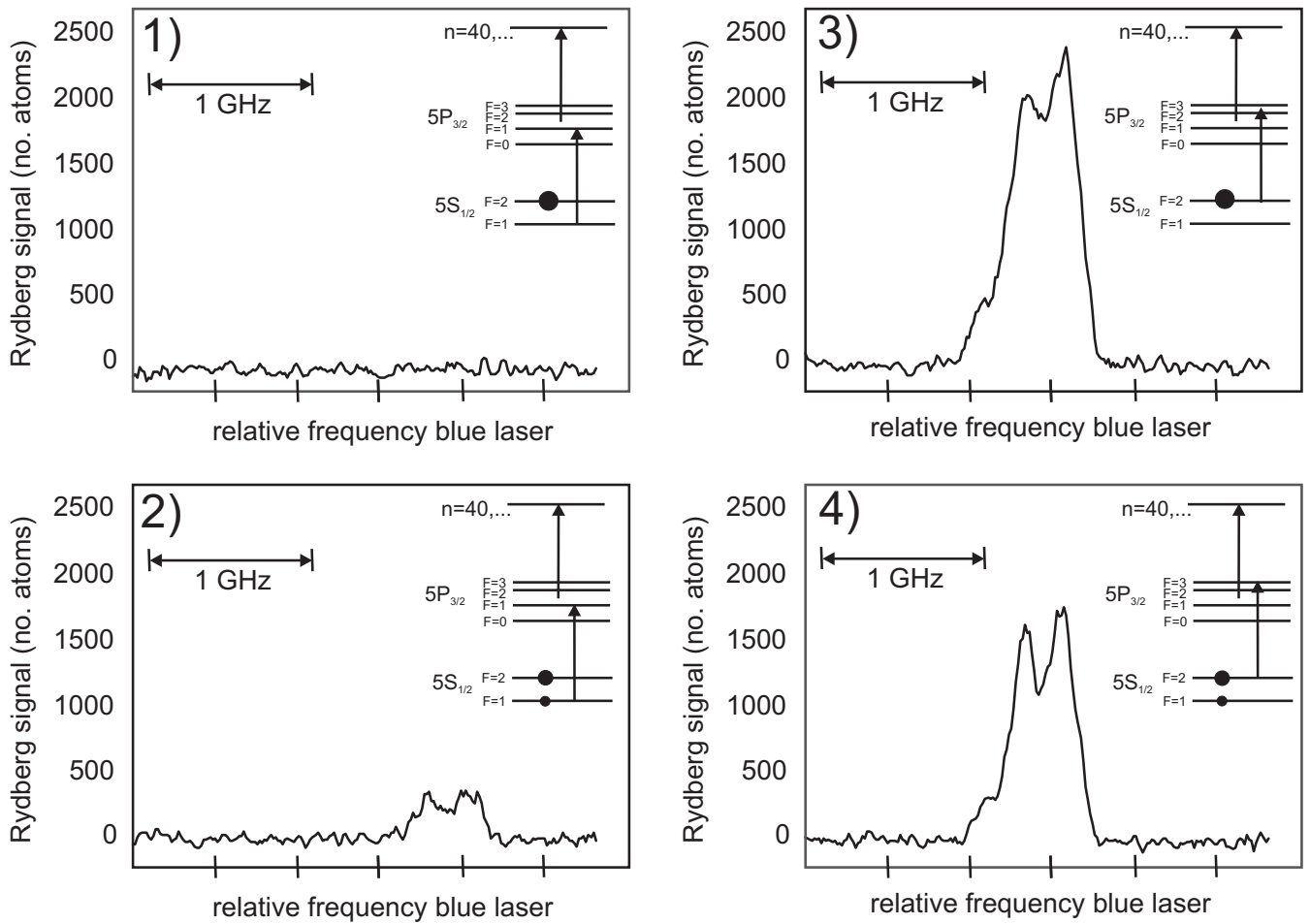}
\caption{State selective excitation of the Rydberg atoms from both
the 5S$_{1/2}$ (F=1) (1 and 2) and (F=2) (3 and 4) states to the
41D duplet. Without optical pumping of the atoms to the F=1 state
no Rydberg excitation is possible (1). With optical pumping before
Rydberg excitation, atoms can be excited to a Rydberg state (2).
Rydberg atom signal without optical pumping before Rydberg
excitation (3) and with optical pumping of the atoms from the F=2
to the F=1 state (4).} \label{fig:6}
\end{figure*}

In the quantum computing scheme with rubidium Rydberg atoms, the
quantum information is stored in both of the hyperfine ground
states (5S$_{1/2}$ F=1, F=2). For this reason, state selectivity
of the ground states in the excitation scheme should be assured.
This means that one of the 5S$_{1/2}$ ground states has to be
accessible for the excitation to the Rydberg state, in a way that
during the excitation, the other state is not affected. To test
this, we excited from both hyperfine ground states to Rydberg
states. In the experiment, we scanned the frequency of the blue
laser across the resonance 5P$_{3/2}$$\rightarrow$ 41D and excited
the atoms on the 5S$_{1/2}$$\rightarrow$5P$_{3/2}$ transition with
the MOT and the repumper laser as red laser system. The repumper
light was resonant with the 5S$_{1/2}$(F=1)$\rightarrow$5P$_{3/2}$
(F=1) transition. We did the excitation with and without optical
pumping of the atoms to the F=1 state before the Rydberg
excitation. In an experiment where no atoms are actively pumped to
the F=1 state, an excitation is only possible from F=2 state to
the Rydberg state (fig. \ref{fig:6},(3)). No atoms can be excited
from the F=1 state to a Rydberg state (fig. \ref{fig:6},(1)). An
excitation from the F=1 state is only possible by actively pumping
the atoms prior to the excitation process in the F=1 state. This
was done by switching off the repumper laser 300 $\mu$s earlier
than the MOT laser before the Rydberg excitation. As shown in fig.
\ref{fig:6} (2) and (4) the number of atoms excited from F=2 to
the n=41D shrinks (4) and an excitation from F=1 to the n=40 D
states becomes possible (2). Since we could not completely switch
off the AOM, we were not able to pump all the atoms in the F=1 and
it was not possible to fully deflate the F=2 state. The measured
frequency shift between the lines is about 460 MHz which is in a
reasonable agreement with the theoretical prediction (423 MHz).
The spectroscopic lines in this experiment are rather broad
because of an electrical field gradient across the atom cloud. To
improve the transfer efficiency, we will use in future experiments
a Raman laser system to transfer the atoms between the hyperfine
ground states.
\section{Autler-Townes Splitting}
\begin{figure*}[h]
\includegraphics[height=7cm]{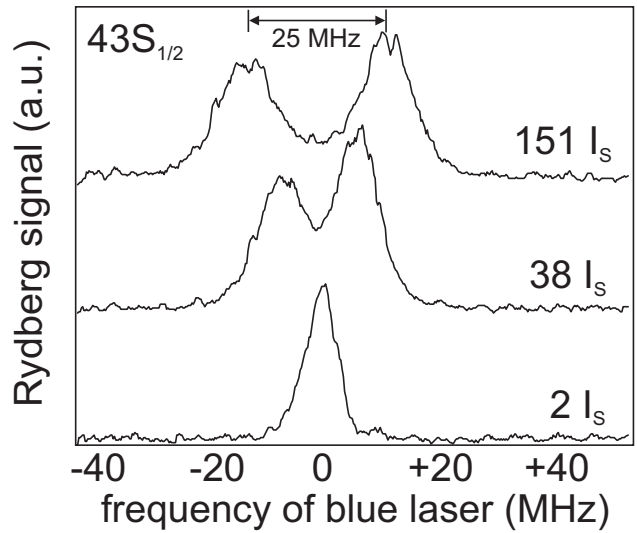}
\caption{Autler-Townes splitting of the 5S$_{1/2}$$\rightarrow$
5P$_{3/2}$ transition probed with a Rydberg transition to the
43S$_{1/2}$ state. For an intensity of 151 I$_{s}$ we get an
Autler-Townes splitting of 25 MHz.} \label{fig:7}
\end{figure*}

In an atomic ensemble that strongly interacts with a coherent
light field, a new set of basis states for the coupled system can
be found \cite{Cohen}. One consequence of this coupling is the
well known Autler-Townes splitting \cite{Autler}. The magnitude of
the splitting is given by the Rabi Frequency of the transition. In
a three level system, it is possible to probe the Autler-Townes
splitting of one state with a transition to an other state and
therefore measure directly the Rabi frequency of the two-state
system \cite{Teo}. The knowledge of the Rabi frequency of our
Rydberg system (for the red and for the blue transition) is
necessary because for coherent excitation to Rydberg states, the
Rabi frequency has to be larger than the inverse timescale of
decoherence processes. To efficiently excite atoms in a two-photon
process to Rydberg state coherently, the Rabi frequencies of the
two transition have to be known and both Rabi frequencies have to
match. We used this method to measure the Rabi Frequency of the
red laser by probing the 5S$_{1/2}$$\rightarrow$5P$_{3/2}$
transition with the transition to the 43S$_{1/2}$ state (see fig.
\ref{fig:7}). By increasing the red laser intensity, the
5S$_{1/2}$$\rightarrow$5P$_{3/2}$ transition splits up into two
lines, where the splitting of the two lines is given by the Rabi
frequency ${\rm \Omega_{r}=c_g \Gamma \sqrt{\frac{I}{2 I_{s}}}}$.
Here $\Gamma$ is the linewidth of the transition, I the intensity
of the laser and I$_{s}$ the saturation intensity of the
transition. c$_g$ accounts for the polarization and state
dependent coupling strength of the different possible transitions,
which is based on the square of the Clebsch-Gordon coefficients.
Since we excite in the MOT with a non-homogeneous magnetic field
and an unpolarized atomic cloud, we have to use here a mean value
for the 5S$_{1/2}$$\rightarrow$ 5P$_{3/2}$ transition of
c$_g=7/15$. As shown in fig. \ref{fig:7}, for a low intensity of 2
I$_{s}$ there is only one single line measured. By increasing the
intensity, the line first broadens up and splits for higher
intensities. From fig. \ref{fig:7} the measured Rabi frequency for
an intensity of 151 I$_{s}$ is 25 MHz. The calculated Rabi
frequency of 24.6 MHz is in good agreement. The reason for the not
identical height of both lines is that the red laser was slightly
detuned with respect to the transition which results in a
different strength of the lines.
\section{Conclusion and Outlook}
In this paper, we described a high resolution system for the
excitation and investigation of $^{87}$Rb atoms in Rydberg states.
All presented experiments were performed on an ensemble of
ultracold atoms in a MOT. With this system, studies in the regime
of a "'frozen Rydberg gas"' are possible. As first experiments, we
tested the resolution and frequency stability of the system by the
investigation of the Stark effect in the vicinity of the n=40
Rydberg states. We could resolve the Stark effect of these states
and even the field dependent splitting of the hyperfine states of
both the 43D$_{3/2}$ and 43D$_{5/2}$ states. The smallest Rydberg
lines we measured had a linewidth of 2.2 MHz which was limited by
the broadening due to the MOT magnetic field.\\
To test the usability for quantum computing approaches, we excited
both atoms spatial and state selectively to Rydberg states. We
furthermore investigated the Autler-Townes splitting of the
5S$_{1/2}$$\rightarrow$ 5P$_{3/2}$ transition which could be
observed by probing this transition with an excitation to a Rydberg state.\\
In the next future, we will be able to start Rydberg excitation
experiments on our $^{87}$Rb Bose-Einstein condensate. Furthermore
an optical lattice is currently set up. So experiments on Rydberg
excitation from atoms trapped inside an optical lattice will be
possible, too.\\[1cm]
{\bf Acknowledgement}
We acknowledge financial support by the 'A8 Quantum Information
Highway' programme of the Landesstiftung Baden-W\"{u}rtemberg\\

\begin{thebibliography}{99}

\bibitem{Gallagher} T.F. Gallagher 1994, {\it Rydberg Atoms} (Cambridge: Cambridge University Press).

\bibitem{Metcalf} H. J. Metcalf, and P. van der Straten, 1999, {\it Laser Cooling and
Trapping} (New York, Springer Verlag).

\bibitem{Anderson} W.R. Anderson, J.R. Veale, and T.F. Gallagher, Phys. Rev. Lett. {\bf 80}, 249 (1998).

\bibitem{Mourachko} I. Mourachko, D. Comparat, F de Tomasi, A. Fioretti, P. Nosbaum, V.M. Akulinm,
and P. Pillet, Phys. Rev. Lett. {\bf 80}, 253 (1998).

\bibitem{Singer} K. Singer, M. Reetz-Lamour, T. Amthor, L. G.
Marcassa, and M. Weidem\"{u}ller, Phys. Rev. Lett. {\bf 93},
163001 (2004).

\bibitem{Tong} D.Tong, S.M. Farooqi, J. Stanojevic, S. Krishnan,
Y.P. Zhang, R. C\^{o}t\'{e}, E.E. Eyler, and P.L. Gould, Phys.
Rev. Lett. {\bf 93}, 063001 (2004).

\bibitem{Carroll} T.J. Carroll, K. Claringbould, A. Goodsell, M.J.
Lim, and M. W. Noel, Phys. Rev. Lett. {\bf 93}, 153001 (2004).

\bibitem{Afrousheh} K. Afrousheh, P. Bohlouli-Zanjani, D. Vagale, A. Mugford, M. Fedorov, and J.D.D. Martin,
 Phys. Rev. Lett. {\bf 93}, 233001 (2004).

\bibitem{Li} Wenhui Li, Paul J. Tanner, and T.F. Gallagher, Phys. Rev. Lett. {\bf
94}, 173001 (2005).

\bibitem{Jaksch} D. Jaksch, J.I. Cirac, P.Zoller, S.L. Rolston, R.
C\^{o}t\'{e}, and M.D. Lukin, Phys. Rev. Lett. {\bf 85}, 2208
(2000).

\bibitem{Lukin} M.D. Lukin, M. Fleischauer, R. C\^{o}t\'{e}, L.M.
Duan, D. Jaksch, J.I. Cirac, and P.Zoller, Phys. Rev. Lett. {\bf
87}, 037901 (2001).

\bibitem{Grimm} See, for example, R. Grimm, M. Weidem\"{u}ller, and Y. B. Ovchinikov,
Adv. At. Mol. Opt. Phys. {\bf 42} 95 (2000).

\bibitem{Grabowski} Axel Grabowski, and Tilman Pfau, Eur.Phys.J.D {\bf 22}, 347-354
(2003).

\bibitem{Teo} B.K. Teo, D. Feldbaum, T. Cubel, J.R. Guest, P.R. Berman, and G. Raithel,
 Phys. Rev. A {\bf 68}, 053407 (2003).

\bibitem{Reichel2} J. Reichel, W. H\"{a}nsel, and T.W.
H\"{a}nsch, Phys. Rev. Lett. {\bf B 83}, 3398, (1999).

\bibitem{Reichel} J. Reichel, W. H\"{a}nsel, P. Hommelhoff, and T.W.
H\"{a}nsch, Appl. Phys. {\bf B 72}, 81, (2001).

\bibitem{Ricci} L. Riccib, M. Weidem\"{u}ller, T. Esslinger, A. Hemmerich,
C. Zimmermann, V. Vuletic, W. K\"{o}nig, and T. W. H\"{a}nsch, Optics
Communications, {\bf 117},541 (1995).

\bibitem{Demtroeder} W. Demtr\"{o}der, 2002, {\it Laser Spectroscopy},
Berlin, Springer Verlag.

\bibitem{Robinson} M.P. Robinson, B. Laburthe Tolra, M.W. Noel,
T.F. Gallagher, and P. Pillet, Phys. Rev. Lett. {\bf 85}, 4466
(2000).

\bibitem{Walz} A. Walz-Flanigan, J.R. Guest, J.-H. Chio, and
G.Raithel, Phys. Rev. A {\bf 69}, 063405 (2004).

\bibitem{Cohen} C. Cohen-Tannoudji, J. Dupont-Roc, and G. Grynberg 1992, {\it Atom-Photon Interactions}
(New York: Wiley-Interscience Publications).

\bibitem{Autler} S.H. Autler, and C.H. Townes, Phys. Rev. {\bf 100}, 703 (1955).

\end{thebibliography}
\end{document}